\documentclass[11pt]{article}
\usepackage{moriond,epsfig}

\def\Journal#1#2#3#4{{#1} {\bf #2}, #3 (#4)}

\def\NPB{{\em Nucl. Phys.} B}
\def\PLB{{\em Phys. Lett.}  B}

\def\PRD{{\em Phys. Rev.} D}

\def\be{\begin{equation}}
\def\ee{\end{equation}}
\def\bea{\begin{eqnarray}}
\def\eea{\end{eqnarray}}

\def\Black{}
\def\Blue{}
\def\Brown{}
\def\slash#1{\setbox0=\hbox{$#1$}#1\hskip-\wd0\dimen0=5pt\advance
\dimen0 by-\ht0\advance\dimen0 by\dp0\lower0.5\dimen0\hbox
to\wd0{\hss\sl/\/\hss}}

\begin{document}

\begin{titlepage}
\null
\vspace{3cm}
\begin{center}
\Large\bf 
\Brown
A Quark-Meson Model for Heavy Mesons:\\
Semi-Leptonic Decays
\Black
\end{center}
\vspace{1.5cm}

\begin{center}
Aldo Deandrea\\
\vspace{0.5cm}
Institut f\"ur Theoretische Physik, Universit\"at Heidelberg,\\
Philosophenweg 19, D-69120 Heidelberg, Germany
\end{center}

\vspace{1.3cm}

\begin{center}
\Brown
{\bf Abstract}\\[0.5cm] \Black
\parbox{14cm}{I consider a model for heavy meson decays based on 
an effective quark-meson Lagrangian. The model is constrained by the known
symmetries of QCD in the $m_Q \to \infty$ limit for the heavy quarks, and 
chiral symmetry in the light quark sector. 
Using a limited number of free parameters it is possible to compute
several phenomenological quantities, e.g. the leptonic $B$ and $B^{**}$ decay
constants; the Isgur-Wise form factors: $\xi$, $\tau (3/2)$,
$\tau (1/2)$, describing the semi-leptonic decays $B \to D^{(*)} l \nu$, 
$B \to D^{**} l \nu$; the form factors for heavy to light decays 
$B \to \rho l \nu$, $B \to a_1 l \nu$. I show that the semileptonic 
heavy-to-light form factors calculated in the model fulfill the general 
relations that hold in QCD in the large energy limit for the final hadron.}
\end{center}

\vspace{2cm}

\begin{center}
\Blue
{\sl To appear in the Proceedings of the XXXIVth Rencontres de Moriond\\
QCD and High Energy Hadronic Interactions\\
Les Arcs 1800, France, March 20--27, 1999.}
\Black
\end{center}

\vfil
\noindent
\Brown
HD-TVP-99-4\\
May 1998
\Black

\end{titlepage}

\newpage
\setcounter{page}{1}

\vspace*{4cm}
\title{A QUARK MODEL FOR HEAVY MESONS:\\
SEMI-LEPTONIC DECAYS}

\author{ALDO DEANDREA}

\address{Institut f\"ur Theoretische Physik, Universit\"at Heidelberg,\\
Philosophenweg 19, D-69120 Heidelberg, Germany}

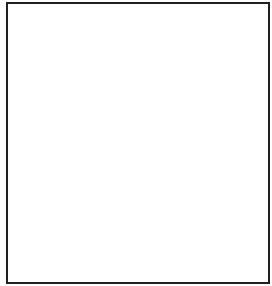
\maketitle\abstracts{I consider a model for heavy meson decays based on 
an effective quark-meson Lagrangian. The model is constrained by the known
symmetries of QCD in the $m_Q \to \infty$ limit for the heavy quarks, and 
chiral symmetry in the light quark sector. 
Using a limited number of free parameters it is possible to compute
several phenomenological quantities, e.g. the leptonic $B$ and $B^{**}$ decay
constants; the Isgur-Wise form factors: $\xi$, $\tau (3/2)$,
$\tau (1/2)$, describing the semi-leptonic decays $B \to D^{(*)} l \nu$, 
$B \to D^{**} l \nu$; the form factors for heavy to light decays 
$B \to \rho l \nu$, $B \to a_1 l \nu$. I show that the semileptonic 
heavy-to-light form factors calculated in the model fulfill the general 
relations that hold in QCD in the large energy limit for the final hadron.}

\section{Introduction}

The model I discuss in this talk conjugates the symmetry approach of 
effective Lagrangians for heavy and light mesons\cite{rep} together with 
dynamical assumptions on chiral symmetry breaking and confinement. 
The reason for adding dynamical assumption is on one side the higher
predictivity one obtains (less free parameters with respect to the
symmetry approach alone) and the possibility of testing the predictions
stemming from these assumptions against experimental data.

The model is also suitable for the description of
higher spin heavy mesons as they can be included in the formalism
in a very easy way \cite{noi}. On the contrary the 
inclusion of higher order corrections, requires
the determination of new free parameters, which proliferate
as new orders are added to the expansion. In this sense the model
allows a simple and intuitive approach to heavy-meson processes
if it is kept at lowest order, while it loses part of its predictive power
if corrections have to be included.

\subsection{Heavy meson field}\label{subsec:hmf}
The lowest negative parity spin doublet $(P,P^*)$ (for charm for instance,
they correspond to $D$ and $D^*$) can be represented by a
$4 \times 4$ Dirac matrix $H$, with one spinor index for the heavy quark
and the other for the light degrees of freedom.

An explicit matrix representation is:
\begin{eqnarray}
H &=& \frac{(1+\slash v)}{2}\; [P_{\mu}^*\gamma^\mu - P \gamma_5 ]\\
{\bar H} &=& \gamma_0 H^{\dagger} \gamma_0~~.
\end{eqnarray}
Here $v$ is the heavy meson velocity, $v^\mu P^*_{a\mu}=0$ and
$M_H=M_P=M_{P^*}$. Moreover $\slash v H=-H \slash v =H$, ${\bar H}
\slash v=-\slash v {\bar H}={\bar H}$ and $P^{*\mu}$ and $P$ are annihilation
operators normalized as follows:
\begin{eqnarray}
\langle 0|P| Q{\bar q} (0^-)\rangle & =&\sqrt{M_H}\\
\langle 0|{P^*}^\mu| Q{\bar q} (1^-)\rangle & = & \epsilon^{\mu}\sqrt{M_H}~~.
\end{eqnarray}
The formalism for higher spin states was introduced by Falk and 
Luke \cite{falk}. In matrix notation, analogous to the one used for the
negative parity states, they are described by
\begin{equation}
S={{1+\slash v}\over 2}[P_{1\mu}^{*\prime} \gamma^\mu\gamma_5-P_0]
\end{equation}
and
\begin{equation}
T^\mu={\frac {1+\slash v}{2}}\left[P_2^{* \mu\nu}\gamma_\nu-\sqrt{\frac 3
2}
P^*_{1\nu}\gamma_5\left(g^{\mu\nu}-\frac 1 3
\gamma^\nu(\gamma^\mu-v^\mu)\right)
\right].
\end{equation}

\subsection{Interaction Lagrangian}\label{subsec:mqi}

The part of the quark-meson effective Lagrangian involving heavy and 
light quarks and heavy mesons is :
\begin{eqnarray}
{\cal L}_{h \ell}&=&{\bar Q}_v i v\cdot \partial Q_v
-\left( {\bar \chi}({\bar H}+{\bar S}+ i{\bar T}_\mu
{D^\mu \over {\Lambda_\chi}})Q_v +h.c.\right)\nonumber \\
&+&\frac{1}{2 G_3} {\mathrm {Tr}}[({\bar H}+{\bar S})(H-S)]
+\frac{1}{2 G_4}
{\mathrm {Tr}} [{\bar T}_\mu T^\mu ] \label{qh1}
\end{eqnarray}
where $Q_v$ is the effective heavy quark field, $\chi$ is the light quark 
field, $G_3$, $G_4$ are coupling constants
and $\Lambda_\chi$ ($= 1$ GeV) has been introduced for dimensional reasons. 
Lagrangian (\ref{qh1}) has heavy spin and flavour symmetry. 

Note that the fields $H$ and $S$ have the same coupling constant. There 
is no symmetry reason for them to be the same.  By putting these two
coupling constant equal, one assumes that this effective quark-meson
Lagrangian can be justified as a remnant of a four quark interaction of
the NJL type by partial bosonization.

The cut-off prescription is also part of the dynamical 
information regarding QCD which is introduced in the model. 
In the infrared the model is not confining and 
its range of validity can not be extended below energies of the order of
$\Lambda_{QCD}$. In practice one introduces an infrared cut-off $\mu$, 
to take this into account.

Models related to the one discussed here, with different regularization
prescriptions and slightly different approach are the one of Bardeen and 
Hill \cite{bardeen} and Holdom and Sutherland \cite{holdom}. The cut-off 
prescription used here is implemented via a proper time regularization. After 
continuation to the Euclidean space it reads, for the light quark propagator:
\begin{equation}
\int d^4 k_E \frac{1}{k_E^2+m^2} \to \int d^4 k_E \int^{1/
\Lambda^2}_{1/\mu^2} ds\; e^{-s (k_E^2+m^2)}\label{cutoff}
\end{equation}
where $\mu$ and $\Lambda$ are infrared and ultraviolet cut-offs.

The cut-off prescription is similar to the one used by Ebert et al. 
\cite{ebert}, with $\Lambda=1.25$ GeV; 
the numerical results are not strongly dependent on the value of $\Lambda$. 
The constituent mass $m$ in the NJL models represents the order
parameter discriminating between the phases of broken and unbroken chiral
symmetry and can be fixed by solving a gap equation, which gives $m$ as a
function of the scale mass $\mu$ for given values of the other parameters. 
Here I take $m=300$ MeV and $\mu=300$ MeV.

\section{Semi-leptonic Decays}

In the following I describe only part of the results obtained using
the quark-meson Lagrangian concerning semi-leptonic decays. More details 
concerning the leptonic decay 
constants and semi-leptonic decays for heavy to heavy mesons can be found 
in \cite{noi}, semi-leptonic decays for heavy to light mesons in \cite{new},
strong decay constants for higher multiplets in \cite{jhep}. A discussion
of the large energy limit of the semileptonic heavy-to-light form
factors not included in previous publications is included at the end.

\subsection{Heavy-to-Heavy Semi-leptonic Decays}\label{subsec:sdff}

As an example of the quantities that can be analytically calculated
in the model, one can examine the Isgur-Wise function $\xi$:
\begin{equation}
\langle D(v^\prime)|{\bar c} \gamma_\mu (1-\gamma_5) b|
B(v)\rangle = \sqrt{M_B M_D} C_{cb}\; \xi(\omega) (v_\mu + v^{\prime }_{\mu})
\end{equation}
where $\omega= v \cdot v^\prime$ and $C_{cb}$
is a coefficient containing
logarithmic corrections depending on $\alpha_s$; within our approximation
it can be put equal to 1: $C_{cb}=1$. At leading order $\xi(1)=1$.
The same universal function $\xi$ also parametrises $B \to D^*$ semileptonic
decay. 

One can compute in a similar way the form factors describing 
the semi-leptonic decays of a meson belonging to the fundamental negative
parity multiplet $H$ into the positive parity mesons in the $S$ and $T$ 
multiplets. Examples of these decays are $B \rightarrow D^{**} l \nu$
where $D^{**}$ can be either a $S$ state or a $T$ state.
These decays are described by two form
factors $\tau_{1/2}, \tau_{3/2}$ \cite{IW2} which
can be computed by a loop calculation similar to the one used to obtain
$\xi(\omega)$. 

The numerical results are reported in Table \ref{tab:3tab}. For a 
comparison with other calculations of these form factors see Morenas 
et al.\cite{pe}. 
\begin{table}
\begin{center}
\caption{Form factors and slopes. $\Delta_H$ in GeV.}\label{tab:3tab}
\vspace{0.2cm}
\begin{tabular}{|c|c|c|c|c|c|c|} 
\hline
\raisebox{0pt}[12pt][6pt]{$\Delta_H$} & 
\raisebox{0pt}[12pt][6pt]{$\xi(1)$} & 
\raisebox{0pt}[12pt][6pt]{$\rho^2_{IW}$} &
\raisebox{0pt}[12pt][6pt]{$\tau_{1/2}(1)$} &
\raisebox{0pt}[12pt][6pt]{$\rho^2_{1/2}$} &
\raisebox{0pt}[12pt][6pt]{$\tau_{3/2}(1)$} &
\raisebox{0pt}[12pt][6pt]{$\rho^2_{3/2}$} \\
\hline
\hline
\raisebox{0pt}[12pt][6pt]{0.3} & 
\raisebox{0pt}[12pt][6pt]{1} & 
\raisebox{0pt}[12pt][6pt]{0.72} & 
\raisebox{0pt}[12pt][6pt]{0.08} & 
\raisebox{0pt}[12pt][6pt]{0.8} &
\raisebox{0pt}[12pt][6pt]{0.48} &
\raisebox{0pt}[12pt][6pt]{1.4} \\
\hline
\raisebox{0pt}[12pt][6pt]{0.4} & 
\raisebox{0pt}[12pt][6pt]{1} & 
\raisebox{0pt}[12pt][6pt]{0.87} & 
\raisebox{0pt}[12pt][6pt]{0.09} & 
\raisebox{0pt}[12pt][6pt]{1.1} &
\raisebox{0pt}[12pt][6pt]{0.56} &
\raisebox{0pt}[12pt][6pt]{2.3} \\
\hline
\raisebox{0pt}[12pt][6pt]{0.5} & 
\raisebox{0pt}[12pt][6pt]{1} & 
\raisebox{0pt}[12pt][6pt]{1.14} & 
\raisebox{0pt}[12pt][6pt]{0.09} & 
\raisebox{0pt}[12pt][6pt]{2.7} &
\raisebox{0pt}[12pt][6pt]{0.67} &
\raisebox{0pt}[12pt][6pt]{3.0} \\
\hline
\end{tabular}
\end{center}
\end{table}
\vspace*{3pt}

An important test of our approach is represented by the
Bjorken sum rule, which states
\begin{equation}
\rho^2_{IW}=\frac{1}{4}+\sum_k \left[|\tau_{1/2}^{(k)}(1)|^2~+~
2|\tau_{3/2}^{(k)}(1)|^2\right]~.
\end{equation}
Numerically we find that the first excited resonances, i.e. the
$S$ and $T$ states ($k=0$) practically saturate
the sum rule for all the three values of $\Delta_H$. From the sum rule 
one can also derive bounds for the slope $\rho^2_{IW}$ of the Isgur-Wise 
function. Neglecting order $\alpha_s$ and $1/m_Q$ corrections the lower 
bound (Bjorken bound \cite{bj}) is $1/4$ while the upper bound (Voloshin 
bound \cite{vol})
is $0.75$. The Bjorken bound is satisfied by our result in table
\ref{tab:3tab}. The Voloshin bound is only marginally satisfied. 
However the Voloshin bound is less stringent as it depends on 
further assumptions \cite{bjvol}.

\subsection{Heavy-to-Light Semi-leptonic Decays}\label{subsec:slff}

The form factors for the semileptonic decays $B\to\rho \ell\nu$ 
can be written as follows ($q=p-p^\prime$):
\begin{eqnarray}
<\rho^+(\epsilon(\lambda),p^\prime)&|&\overline{u}\gamma_\mu 
(1-\gamma_5)b|\bar{B^0}(p)>
= \frac{2 V(q^2)}{m_B+m_{\rho}}\epsilon_{\mu\nu\alpha\beta}
\epsilon^{*\nu}p^\alpha p^{\prime\beta}\nonumber \\
&-& i \epsilon^{*}_{\mu}(m_{B} + m_{\rho})  A_{1}(q^{2})
+ i (\epsilon^{*}\cdot q)
\frac{(p + p^\prime)_{\mu}}{m_B +  m_{\rho}}  A_{2}(q^{2})
\\
&+& i  (\epsilon^{*}\cdot  q)
\frac{2  m_{\rho}}{q^{2}} q_{\mu} [A_{3}(q^{2})  - A_{0}(q^{2})]
\nonumber\;\; ,
\end{eqnarray}
where
\begin{equation}
A_{3}(q^{2})  = \frac{m_{B} + m_{\rho}}{2  m_{\rho}} A_{1}(q^{2})
- \frac{m_{B} - m_{\rho}}{2  m_{\rho}} A_{2}(q^{2}) \;\; ,
\end{equation}
The calculation of these form factors in the model arises from two sources:
``direct'' diagrams in which the weak current couples directly to the quarks
in the light and heavy mesons, and ``polar'' contributions where the weak
current is coupled to the heavy and light mesons by an intermediate meson
state. The expressions for the form factors are quite lengthy and can be 
found in \cite{new}. The numerical results are in Table \ref{t:tab2}.
\begin{table}
\hfil
\vbox{\offinterlineskip
\halign{&#& \strut\quad#\hfil\quad\cr
\hline
& && CQM \cite{new} && Potential Model \cite{ladisa} 
&& SR \cite{ballcol} && Latt. + LCSR \cite{lattice} &\cr
\hline
& $V^{\rho}(0)$ && $-0.01 \pm 0.25$ && $0.45~\pm~0.11$ && 
$0.6~\pm~0.2$ && $0.35^{+0.06}_{-0.05}$ &\cr
& $A_1^{\rho}(0)$ && $0.58\pm 0.10$ && $0.27~\pm~0.06$ &&
$0.5~\pm~0.1$ && $0.27^{+0.05}_{-0.04}$ &\cr
& $A_2^{\rho}(0)$ && $0.66\pm 0.12$ && $ 0.26~\pm~0.05$ && 
$0.4~\pm~0.2$ && $0.26^{+0.05}_{-0.03}$ &\cr
& $A_0^{\rho}(0)$ && $0.33 \pm 0.05$ && $0.29~\pm~0.09$ && $ 0.24\pm 0.02$ 
&& $0.30^{+0.06}_{-0.04}$ &\cr
\hline}}
\caption{Form factors for the transition $B \to \rho$ at $q^2=0$. The results
of CQM are compared with the outcome of other theoretical calculations:
potential models, QCD sum rules (SR), calculations
involving both lattice and light cone sum rules. The large error of 
$V^\rho (0)$ in our approach is due to the large cancellation between the 
direct and polar contribution.}
\label{t:tab2}
\end{table}
For the $B\to \rho \ell\nu$ decay width and branching ratio
the model predicts (using $V_{ub}=0.0032$, 
$\tau_B=1.56 \times 10^{-12}$ s):
\begin{eqnarray}
{\cal B}(\bar B^0 \to \rho^+ \ell \nu) &=& (2.5 \pm 0.8) \times 10^{-4} \nonumber \\
\Gamma_0(\bar B^0 \to \rho^+ \ell \nu) &=& (4.4 \pm 1.3) \times 10^{7} \; s^{-1} 
\nonumber \\
\Gamma_+ (\bar B^0 \to \rho^+ \ell \nu)&=& (7.1 \pm 4.5) \times 10^{7} \; 
s^{-1} \nonumber \\
\Gamma_- (\bar B^0 \to \rho^+ \ell \nu)&=& (5.5 \pm 3.7) \times 10^{7} \; 
s^{-1} \nonumber \\
(\Gamma_+ + \Gamma_-) (\bar B^0 \to \rho^+ \ell \nu)&=& (1.26 \pm 0.38) \times
10^8 \; s^{-1}
\end{eqnarray}
where $\Gamma_0$, $\Gamma_+$, $\Gamma_-$ refer to the $\rho$ helicities.
This decay was observed by the CLEO collaboration \cite{cleo}:
\begin{equation}
{\cal B}(B^0 \to \rho^- \ell^+ \nu)=(2.5 \pm 0.4^{+0.5}_{-0.7}\pm0.5) \;
\times \; 10^{-4} \;.
\end{equation}
in good agreement with what is predicted by the constituent quark model.

\subsection{Heavy-to-Light Form Factors and Final Hadron Large Energy Limit}

It is interesting to examine a particular limit for the $B \to \rho$ 
semileptonic form factors,
namely the one of heavy mass for the initial meson and of large energy 
for the final one. In this limit the expressions of the form factors
simplify and for $B \to V l \nu$, where $V$ is the $\rho$ in the following
example, they reduce only to two independent functions (see J. Charles et 
al.\cite{leet} for details). The four-momentum of 
the heavy meson is written as $p= M_H v$ in terms of the mass and the 
velocity of the heavy meson. The four-momentum of the $\rho$ is written as
$p'=E n$ where $E=v \cdot p'$ is the energy of the light meson and $n$ is
a four-vector defined by $v \cdot n=1, n^2=0$. This peculiar large energy 
limit is defined as :
\be
\Lambda_{QCD}, m_V << M_H, E
\ee
keeping $v$ and $n$ fixed and $m_V$ is in our example the mass of the 
$\rho$. In agreement with J. Charles et al.\cite{leet} I find the following 
result :
\bea
A_0(q^2)&=&\left(1-\frac{m_V^2}{M_H E}\right)\zeta_{||}(M_H,E)
+\frac{m_V}{M_H}\,\zeta_{\perp}(M_H,E)\\
A_1(q^2)&=&\frac{2E}{M_H + m_V}\,\zeta_{\perp}(M_H,E) \label{fatta}\\
A_2(q^2)&=&\left(1+\frac{m_V}{M_H}\right)\left[\zeta_{\perp}(M_H,E)-
\frac{m_V}{E}\zeta_{||}(M_H,E)\right]\\
V(q^2)&=&\left(1+\frac{m_V}{M_H}\right)\zeta_{\perp}(M_H,E).
\label{fattv}
\eea
{}From (\ref{fattv}) and (\ref{fatta}) one can immediately obtain the 
relation:
\be
\frac{V(q^2)}{A_1(q^2)}=\frac{(M_H+m_V)^2}{M_H^2+m_V^2-q^2}
\ee
The explicit expressions for $\zeta_{||}$ and $\zeta_{\perp}$
are as follows in the constituent quark model:
\bea
\zeta_{||}(M_H,E)&=&\frac{\sqrt{M_H Z_H}\; m_V^2}{2 E f_V} \Big[
I_3\left(\frac{m_V}{2}\right)-I_3\left(-\frac{m_V}{2}\right)
\nonumber \\
&+&4 \Delta_H m_V\; Z(\Delta_H)
\Big] \sim \frac{\sqrt{M_H}}{E} \eea
\be
\zeta_{\perp}(M_H,E)=\frac{\sqrt{M_H Z_H}\; m_V^2} {2 E f_V}
\left[I_3(\Delta_H) + m_V^2 \; Z(\Delta_H) \right]
\sim \frac{\sqrt{M_H}}{E}
\label{rellt}
\ee
where terms proportional to the constituent light quark mass $m$ have been
neglected. $Z_H$ is the renormalization constant for the heavy meson 
\cite{noi}. $\Delta_H$ is the difference between the heavy meson and the
heavy quark mass. This quantity stays finite in the limit. The full
expressions for the form factors also involve $\Delta'=\Delta_H - E$.
However using the asymptotic expansion for the error function entering in 
the integrals one can show that in the limit $E \to \infty$ these terms
vanish. The functions appearing in the previous expression are: 
\bea
I_3(\Delta) &=& - \frac{iN_c}{16\pi^4} \int^{\mathrm {reg}}
\frac{d^4k}{(k^2-m^2)(v\cdot k + \Delta + i\epsilon)}\nonumber \\ &=&{N_c
\over {16\,{{\pi }^{{3/2}}}}} \int_{1/{{\Lambda}^2}}^{1/{{\mu }^2}} {ds \over
{s^{3/2}}} \; e^{- s( {m^2} - {{\Delta }^2} ) }\; \left( 1 + {\mathrm {erf}}
(\Delta\sqrt{s}) \right) \; ,
\eea
where erf is the error function and
\begin{equation}
Z(\Delta) =  \frac{iN_c}{16\pi^4} \int^{\mathrm {reg}}
\frac{d^4k}{(k^2-m^2)[(k+q)^2-m^2](v\cdot k + \Delta + i\epsilon)}\; .
\end{equation}
The fact that the model fulfills the large energy limit was not obvious
from the start. This is a further test of consistency of the model. 
Concerning the scaling properties of $\zeta_{||}$ and 
$\zeta_{\perp}$, the asymptotic $E$-dependence is not predicted by the
large energy limit. As $E \sim M$ at $q^2=0$ the Feynman mechanism 
contribution to the form factors would indicate a $1/E^2$
behaviour rather than the $1/E$ found in the model. Note however that
the $E$-dependence is not rigorously established in QCD.

\section{Conclusions}

{}From an effective Lagrangian at the level of mesons and constituent
quarks, it is possible to compute meson transition amplitudes by 
evaluating loops of heavy and light quarks.  
The agreement with data, when available, is good. 
The model is able to describe a number of essential features of 
heavy meson physics in a simple and compact way, in particular Isgur-Wise
scaling in the heavy-to-heavy semileptonic decays and the large energy limit
for the heavy-to-light ones.

\section*{Acknowledgments}
I acknowledge the support of the EC-TMR (European Community 
Training and Mobility of Researchers) Program on ``Hadronic Physics with 
High Energy Electromagnetic Probes''. I would like to thank
J. Charles, A. Le Yaouanc, A.D. Polosa and O. P\`ene for discussion and 
G. Korchemsky for invitation at LPTHE Orsay.


\section*{References}
\begin{thebibliography}{99}

\bibitem{rep}
R. Casalbuoni et al., {\em Phys. Rep.} {\bf 281} 145 (1997), 
{\tt hep-ph/9605342}.

\bibitem{noi}
A. Deandrea, N. Di Bartolomeo, R. Gatto, G. Nardulli, A.D. Polosa,
\Journal{\PRD}{58}{034004}{1998}, {\tt hep-ph/9802308}.

\bibitem{falk}
A. Falk and M. Luke \Journal{\PLB}{292}{119}{1992}, {\tt hep-ph/9206241}.

\bibitem{bardeen}
W. H. Bardeen and C. T. Hill, \Journal{\PRD}{49}{409}{1994},
{\tt hep-ph/9304265}.

\bibitem{holdom}
B. Holdom and M. Sutherland, \Journal{\PRD}{47}{5067}{1993},
{\tt hep-ph/9211226}.

\bibitem{ebert}
D. Ebert, T. Feldmann, R. Friedrich and H. Reinhardt,
\Journal{\NPB}{434}{619}{1995}, {\tt hep-ph/9406220}; 
D. Ebert, T.~Feldmann and H.~Reinhardt,
\Journal{\PLB}{388}{154}{1996}, {\tt hep-ph/9608223}.

\bibitem{new}
A. Deandrea, R. Gatto, G. Nardulli, A.D. Polosa, \Journal{\PRD}{59}{074012}
{1999}, {\tt hep-ph/9811259}.

\bibitem{jhep}
A. Deandrea, R. Gatto, G. Nardulli, A.D. Polosa, {\em JHEP}
{\bf 02} 021 (1999), {\tt hep-ph/9901266}.

\bibitem{IW2}
N. Isgur and M. B. Wise, \Journal{\PRD}{43}{819}{1991}.

\bibitem{pe}
V. Morenas et al., \Journal{\PRD}{56}{5668}{1997}, {\tt hep-ph/9706265}.

\bibitem{bj}
J.D. Bjorken, in Proceedings of the 4th Rencontre de la Valle d'Aoste,
La Thuile, Italy, 1990, ed. M. Greco (Editions Frontieres, Gif-sur-Yvette,
France, 1991).

\bibitem{vol}
M.B. Voloshin, \Journal{\PRD}{46}{3062}{1992}.

\bibitem{bjvol}
C.G. Boyd, Z. Ligeti, I.Z. Rothstein, M.B. Wise, 
\Journal{\PRD}{55}{3027}{1997}, {\tt hep-ph/9610518}.

\bibitem{ladisa}
P. Colangelo, F. De Fazio, M. Ladisa, G. Nardulli, P. Santorelli and 
A. Tricarico, {\em Eur.Phys.J.}{\bf C8} 81 (1999), {\tt hep-ph/9809372}.

\bibitem{ballcol}
P. Ball,\Journal{\PRD}{48}{3190}{1993}, {\tt hep-ph/9305267}; 
P. Colangelo, F. De Fazio and P. Santorelli, \Journal{\PRD}{51}{2237}{1995},
{\tt hep-ph/9409438}; P. Colangelo, F. De Fazio, P. Santorelli and 
E. Scrimieri, \Journal{\PRD}{53}{3672} {1996}, {\tt hep-ph/9510403}; 
{\em ibid.} \Journal{\PRD}{57}{3186}{1998} (E).

\bibitem{lattice}
L. Del Debbio {\it et al.}, (UKQCD Collaboration),
\Journal{\PLB}{416}{392}{1998}, {\tt hep-lat/9708008}.

\bibitem{cleo}
CLEO Collab., J. P. Alexander et al., Phys. Rev. Lett. {\bf 77} (1996) 5000.

\bibitem{leet} 
J. Charles et al. hep-ph/9812358 to appear in {\em Phys. Rev. D};
{\em ibid.} \Journal{\PLB}{451}{187}{1999}, {\tt hep-ph/9901378}. 

\end{thebibliography}
\end{document}